\documentclass[aps,prl,showpacs,twocolumn,superscriptaddress]{revtex4}
\usepackage{amsmath}
\usepackage{amssymb}
\usepackage{graphicx}
\begin{document}

\title{Flux state and anomalous quantum Hall effect in square Kondo lattice}
\author{Xiao Chen}
\affiliation{Laboratory of Solid State Microstructures, Nanjing University, Nanjing 210093, China}
\author{Shuai Dong}
\affiliation{Laboratory of Solid State Microstructures, Nanjing University, Nanjing 210093, China}
\author{J.-M. Liu}
\affiliation{Laboratory of Solid State Microstructures, Nanjing University, Nanjing 210093, China}
\affiliation{International Center for Materials Physics, Chinese Academy of Sciences, Shenyang 110016, China}
\date{\today}

\begin{abstract}
The anomalous Hall effect (AHE) around the flux state in square
Kondo lattice is investigated. By introducing the lattice distortion
and local chirality, the square Kondo lattice can break the parity
symmetry and time reversal symmetry spontaneously, and thus generate
a topological nontriviality in the band structure associated with
the AHE. Moreover, a possible realization of this AHE in
multiferroic TbMnO$_3$ is discussed.
\end{abstract}
 \maketitle

In condensed matters, novel spin orders often lead to novel physical
phenomena. For instance, some multiferroics have a spiral spin order
which breaks the spatial inversion symmetry and gives rise to
ferroelectric polarization, which's origin is completely different
from conventional ferroelectricity \cite{Kimura:Rmr,Wang:Ap}. This
spiral spin order can be scaled by a vector spin chirality (VSC)
$\textbf{S}_{i}\times\textbf{S}_{j}$ and has become an important
concept in the physics of spin current and spin liquid
\cite{Katsura:Prl,Park:Prl}. Besides the VSC, there is another
scalar spin chirality defined by:
\begin{equation}
\chi_{ijk}=\textbf{S}_i\cdot(\textbf{S}_j\times\textbf{S}_k),
\end{equation}
which breaks the parity (P) and time reversal (T) symmetry and was
first proposed by Wen \textit{et al} \cite{Wen:Prb}. It is clear
that chirality $\chi$ would be nonzero for a noncoplanar spin order.

On the other hand, recent studies showed that the noncoplanar spin
order is relevant to the intrinsic anomalous Hall effect (AHE)
observed in Kondo lattice system, in which the noncoplanar
background spin texture acts as a gauge field for itinerant
electrons propagating in the lattice
\cite{Ohgushi:Prb,Shindou:Prl,Neubauer:Prl,Lee:Prl,Yi:arxiv,Taguchi:arxiv,Nagaosa:arxiv,Martin:Prl,Onoda:Prl,Taguchi:Science}.
Generally speaking, the intrinsic AHE has the topological origin and
can be characterized by the Berry phase and Chern number. To
manifest this mechanism, the system should break the P-symmetry and
T-symmetry spontaneously and simultaneously
\cite{Onoda:Jpsj,Nagaosa:arxiv}. For a Kondo lattice model, the
T-symmetry can be violated for those spin configurations with the
local spin chirality, while the P-symmetry can be broken in some
geometrically frustrated lattices. For instance, Ohgushi \textit{et
al} once discussed the AHE on Kagome lattice where a finite local
spin chirality in the three-site unit cell can generate nonzero Hall
conductance \cite{Ohgushi:Prb}. This model has been extended to
other geometrically frustrated systems such as the three-dimensional
($3$D) pyrochlore lattice
\cite{Shindou:Prl,Martin:Prl,Taguchi:Science}. Therefore, a
geometrically frustrated lattice with spin chirality would be of
significance in terms of AHE physics.

Unfortunately, the square lattice usually has no geometrically
frustrated structure, and thus it is not easy to violate the
P-symmetry since the chiralities on adjacent plaquettes tend to
cancel each other due to the lattice symmetry and thus the AHE
becomes hard to realize in the square lattice. However, the square
lattice (and its distorted forms)  takes up the majority in the
practical Kondo lattice materials such as colossal
magnetroresistance manganites \cite{Dagotto:Pr} and Fe-based
pnictide superconductors \cite{Kamihara:Jacs}. Besides, the
realization of AHE on the square lattice is also fundamentally
important and physically interesting \cite{Goldman:Pra,
Goldman:Prl}.

In this paper, the AHE on the Kondo square lattice is realized
theoretically by introducing two mechanisms to break the P-symmetry.
One is to induce some lattice distortions which can lead to the
change of hopping amplitude of itinerant electrons. The other is to
construct a special unit cell which breaks the P-symmetry. These two
mechanisms are different from the previous considered spin-orbit
interaction (SOI) which can directly generate topological
nontrivality in the band structure and associate with the AHE
\cite{Onoda:Jpsj,Nagaosa:arxiv,Takahashi:Prl}. Our discussion will
be primarily restricted to a 'flux' state at the half filling of
one-band Kondo model. We hope that these two mechanisms can be
alternating approaches (other than the SOI mechanism) to the AHE in
the square lattice, and eventually realized in some real materials.

The Hamiltonian of one-band Kondo lattice model on the two-dimensional ($2$D) square lattice can be written as:
\cite{Dagotto:Pr}
\begin{eqnarray}
\nonumber
H&=&-\sum_{NN}t_1c_{i,\sigma}^{\dag}c_{j,\sigma}-\sum_{NNN}t_2c_{i,\sigma}^{\dag}c_{k,\sigma}\\
\nonumber &&-J_{\rm H}\sum_i\textbf{S}_i\cdot c_{i,\alpha}^{\dag}\sigma_{\alpha\beta}c_{i,\beta}\\
&&+J_1\sum_{NN}\textbf{S}_i\cdot\textbf{S}_j+J_2\sum_{NNN}\textbf{S}_i\cdot\textbf{S}_j,
\end{eqnarray}
where the first term describes the electron hopping between the
nearest-neighbor (NN) sites, and the hopping amplitude $t_1$ is
taken as the energy unit. The second term is the electron hopping
between the next-nearest-neighbor (NNN) sites. In the following,
$t_2$ is arbitrarily set as $0.25$ as an example, since the AHE
result is qualitatively independent of its exact value as long as it
is nonzero, which will be further dicussed below. The third term is
the Hund coupling linking the itinerant electrons with the
background spins $\textbf{S}$ (assumed classical and normalized as
$|\textbf{S}|=1$) where $J_{\rm H}$ is the coupling factor. The last
two terms are the antiferromagnetic (AFM) superexchanges between the
background spins with $J_1$ and $J_2$ as the coefficients for the NN
and NNN sites respectively. This model has been extensively
investigated for various transitional metals' oxides, and more
details of this model can be found in Ref.~\onlinecite{Dagotto:Pr}.

For the third term, by applying a canonical transformation, the Hamiltonian can be simplified by adopting the site-dependent spin-polarization axis. In the $J_{\rm H}\rightarrow\infty $ limit, the spin of the hopping electron is forced to align parallel to the on-site $\textbf{S}$, and thus the hopping terms in the Hamiltonian can be transferred into the form $t_{ij}^{eff}c_{i}^{\dag}c_{j}$, with the effective hopping integral:
\begin{eqnarray}
\nonumber
t_{ij}^{eff}&=&t[\cos\frac{\theta_i}{2}\cos\frac{\theta_j}{2}+\sin\frac{\theta_i}{2}\sin\frac{\theta_j}{2}e^{-i(\varphi_i-\varphi_j)}]\\
&=& te^{i\varphi_{ij}}\cos\frac{\theta_{ij}}{2},
\end{eqnarray}
where $t$ can be $t_1$ or $t_2$. $\theta$ and $\varphi$ are the
polar coordinates of spin $\textbf{S}$. The phase factor
$\varphi_{ij}$ can be viewed as the gauge vector potential and
$\theta_{ij}$ is the angle between $\textbf{S}_i$ and $\textbf{S}_j$
\cite{Ohgushi:Prb}. When the itinerant electrons move along a closed
loop, they can feel the induced gauge flux which is
indistinguishable from the magnetic flux. This gauge flux is related
to the spin chirality and leads to the AHE
\cite{Ohgushi:Prb,Nagaosa:arxiv,Onoda:Prl,Taguchi:Science}.

For this model, the groud state can be calculated with the
variational method, namely, by comparing the ground state energies
of several preset spin configurations . Around half filling, when
the $J_1$=$J_2>0.15$, the ground state is the flux state. This state
has four sites in the unit cell, as shown in Fig.~1(a). In one
plaquette, the neighboring background spins are perpendicular to
each other. When an itinerant electron travels around the plaquette,
it can acquire an additional $\pi$ flux. In fact, this flux state
was reported earlier in some similar systems
\cite{Yamanaka:Prl,Agterberg:Prb}. The flux phase can have many
degenerate states by shifting and rotating the spin structures.
Thus, a specific state, with ($\textbf{S}_1$, $\textbf{S}_3$) along
the $z$-axis and ($\textbf{S}_2$, $\textbf{S}_4$) along the
$x$-axis, will be adopted in the following study, which forms a spin
order in the $x$-$z$ plane. In practical calculation, to remove the
degeneration and stabilize this state, a small anisotropic energy
$H_2=\sum K_{\alpha}{S_{\alpha}^z}^2$ (subscript $\alpha$=$1$-$4$,
$K_1$=$K_3$=$-0.02$, $K_2$=$K_4$=$0.02$) is also considered, which
favors ($\textbf{S}_1$, $\textbf{S}_3$) along the $z$-axis and
($\textbf{S}_2$, $\textbf{S}_4$) in the $x$-$y$ plane. In fact, the
flux state is with coplanar spin order and the AHE conductance is
forbidden. Therefore, additional contributions should be included
for allowing the AHE.

In real materials, due to the ionic size mismatch or competing
exchange interactions, the square lattice would be distorted more or
less. For instance, in multiferroic RMnO$_3$, the Mn-O-Mn chain is
distorted in noncentrosymmetric manners caused by the
Dzyaloshinsky-Moriya (DM) interaction, giving rise to the staggered
Mn-O-Mn angles \cite{Kimura:Rmr,Arima:Prl}. For simplicity, a simple
lattice distortion  mode is adopted, as shown in Fig.~1(b), where
the first and third cations are displaced from the original
positions along the opposite directions. Due to the $x$-$y$
symmetry, the cation displacement along $x$ and $y$ are assumed to
be the same. In a first order approximation, the corresponding
hopping amplitude varies linearly with the tiny lattice distortion,
and thus the NN hopping amplitude $t_1'=t_1\pm d$ where $d$ is the
tiny amendment caused by the distortion. The influence to the NNN
hopping amplitude $t_2$ is not considered because the exact value of
NNN hopping is not qualitatively important to obtain the AHE . Take
one chain of the lattice for example, the hopping amplitude becomes
staggeringly ordered and the P-symmetry of the lattice is broken.

\begin{figure}
\vskip -0.6cm \centerline{\includegraphics[width=9cm]{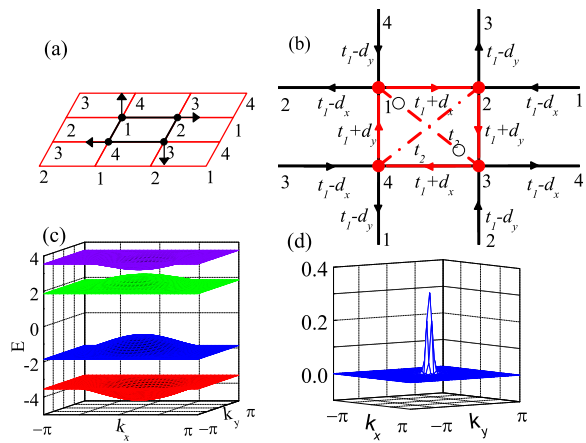}}
\vskip -0.7cm \caption{(Color online) (a) The flux state with the
four-site unit cell. (b) The lattice is distorted with atoms $1$ and
$3$ moving in reversed direction with the new positions shown as
open circles. The NN hopping amplitude $t_1$ is varied
correspondingly. The arrows on bonds indicate the signs of the
phases of the $t_{ij}^{eff}$. (c) The band structure of Eq.~(5) with
the parameters sets $t_{r1}=1$, $t_{r2}=0.25$, $d=1$ and
$\phi=\pi/6$. (d) Gauge flux density of the $3$-th band of (c).}
\vskip -0.5cm
\end{figure}

With this P-symmetry broken configuration, the spin order may be not
exactly confined on the $x$-$z$ plane, i.e., spins $\textbf{S}_1$
and $\textbf{S}_3$ tilt slightly from the $z$-axis and are not
parallel to each other. Therefore, the local spin order may become
non-coplanar, with each plaquette having a gauge flux penetrating
it. In this condition, the NN hopping can be approximately modified
to the following form \cite{Ohgushi:Prb}:
\begin{equation}
t_{ij}^{eff}=t_{r1}e^{i\varphi_{ij}}
\end{equation}
where $t_{r1}$ is the renormalized NN hopping amplitude. We take
$\varphi_{ij}=\phi(-\phi)$ for the hopping direction along (opposite
to) the arrow direction as shown in Fig.~2(b). In addition, the
renormalized NNN hopping amplitude is set as a real constant
$t_{r2}$ for simplicity. The Hamiltonian matrix for this model can
be written in the momentum space:
\begin{eqnarray}
H(k)=\left(
  \begin{array}{cccc}
    0 & e^{i\phi}f_1 & f_3 & e^{-i\phi}f_2 \\
    e^{-i\phi}f_1^{\ast} & 0 & e^{i\phi}f_2 & f_3 \\
    f_3 & e^{-i\phi}f_2^{\ast} & 0 & e^{i\phi}f_1^{\ast} \\
    e^{i\phi}f_2^{\ast} & f_3 & e^{-i\phi}f_1 & 0 \\
  \end{array}
\right)
\end{eqnarray}
where $f_1(k)$=$t_{r1} \cos(k_x/2)$+$id \sin(k_x/2)$,
$f_2(k)$=$t_{r1} \cos(k_y/2)$+$id \sin(k_y/2)$,
$f_3(k)$=$t_{r2}[\cos((k_x-k_y)/2)$+$\cos((k_x+k_y)/2)]$. Now, the
AHE conductance can be calculated. The contribution to the AHE
conductance from each band is written as
\cite{Onoda:Jpsj,Ohgushi:Prb}:
\begin{eqnarray}
\nonumber \sigma_{xy}^n&=&\frac{e^2}{h}\frac{1}{2\pi i}\int_{BZ}
d^2k\textbf{z}\cdot\nabla_k\times
\textbf{A}_n(k)\\
\nonumber &=&\frac{e^2}{h}\frac{1}{2\pi i}\int_{BZ}
d^2k\textbf{z}\cdot\textbf{B}_n(k)\\
&=&\frac{e^2}{h}C_n,
\end{eqnarray}
where $\textbf{A}_n(\textbf{k})=<nk|\nabla|nk>$ is the vector
potential defined with the $n$-th wave function,
$\textbf{B}_n(\textbf{k})$ is the gauge flux density and $C$ is the
so-called first Chern number. At $d=0$ case, namely, the ideal
square lattice without any distortion, the P-symmetry is maintained,
leading to zero Chern number for each band. At $d\neq0$ case, the
P-symmetry is broken. The calculation indicates that the Chern
number for each band is $C$=$[0, 0, 1, -1]$ at $\phi<\pi/4$, and
$C$=$[1, -1, 0, 0]$ at $\phi>\pi/4$ (at $\phi=\pi/4$, the T-symmetry
is conserved, corresponding to the flux state). For instance, at
$\phi=\pi/6$, the band structure is shown in Fig.~1(c), and the
gauge flux density $\textbf{B}(\textbf{k})$ of the $3$-th band is
shown in Fig.~1(d).

For Hamiltonian Eq.~(2), the nonzero NNN hopping term is essentially
important to generate the topological nontriviality in the band
structure. This can be intuitively understood as follows. For
Eq.~(5), in the large $d$-limit, $f_1(k)\approx id\sin(k_x/2)$,
under the unitary transformation $UHU^T\rightarrow{H'}$
\cite{supplement:none}, the Hamiltonian matrix Eq.~(5) can be
further decoupled into the form:
\begin{eqnarray}
{H'}=\left(
  \begin{array}{cc}
    h_1(k) & 0 \\
    0 & h_2(k) \\
  \end{array}
\right)
\end{eqnarray}
where $h_1(k)$ and $h_2(k)$ are both $2\times 2$ matrixes.

For the matrix $h_1(k)$, we have:
\begin{eqnarray}
h_1(k)=\left(
  \begin{array}{cc}
    d\sin\phi f_4 & -id\cos\phi f_5-f_3 \\
    id\cos\phi f_5-f_3 & -d\sin{\phi} f_4 \\
  \end{array}
\right)
\end{eqnarray}
where $f_4(k)=\sin(k_x/2)+\sin(k_y/2)$,
$f_5(k)=\sin(k_x/2)-\sin(k_y/2)$. Around $k=(\pi,\pi)$, the electron
can be considered as a generalized Dirac fermion and the effective
Hamiltonian $h_1(k)$ ($h_2(k)$ can be treated in a similar way) is
\begin{equation}
h_1(k)=-\frac{t_{r2}}{2}k_x'k_y'\sigma^x+\frac{d\cos{\theta}}{8}(k_y'^2-k_x'^2)\sigma^y+2d\sin{\phi}\sigma^z,
\end{equation}
where $k_x'\equiv k_x-\pi$, $k_y'\equiv k_y-\pi$. The general form
of Eq.~(10) was thoroughly addressed in
Ref.~\onlinecite{Onoda:Jpsj}, and the corresponding Chern number for
the upper and lower bands is $C=\pm 2sgn(t_{r2}/d)$ (at
$\phi=\pi/6$) where $sgn$ is the sign operator. If $t_{r2}=0$, $C=0$
for both bands, indicating that the NNN hopping term is
indispensable in generating the Hall conductance.

Consequently, one can argue that the lattice distortion provides an
effective method to break the P-symmetry, and yet the non-coplanar
spin order as the ground state is absolutely necessary. In fact, the
non-coplanar spin order can also be accomplished by further taking
into account the frustrated magnetic interaction. For instance, by
adding the third neighbor (3rd N) superexchange interaction
$H_3=\sum_{3rd N} J_3\textbf{S}_i\cdot\textbf{S}_j$ to the distorted
Kondo lattice model described by Eq.~(2), the original flux state at
the half filling evolves into a state in Fig.~2(a). In this
condition, the unit cell expands to eight-site and is formed with
two interlaced square sublattices, compared with the original
four-site unit cell (formed by $\textbf{S}_6$, $\textbf{S}_3$,
$\textbf{S}_7$ and $\textbf{S}_4$). For this spin configuration, we
can straightly calculate the Hall conductivity at zero temperature
by using the Kubo formula \cite{Onoda:Jpsj}:
\begin{equation}
\sigma_{xy}=\frac{e^2}{h}\frac{1}{2\pi i}\sum_{nmk}
\frac{<nk|J_x|mk><mk|J_y|nk>-h.c.}{[\varepsilon_n(k)-\varepsilon_m(k)]^2},
\end{equation}
where $|nk>$ is the occupied state and $|mk>$ is the empty state.
$\varepsilon$ denotes the eigen-energy. $J_x$ ($J_y$) is the $x$
($y$) components of the current operator. The calculation is
performed in momentum space, with the summation over all the eight
bands. For example, with a finite $J_3$ (see Fig. 2(a)'s caption),
$\sigma_{xy}$=$0.13$ e$^2$/h at half filling, clearly indicating
that this ground state does exhibit the AHE. Besides, the Hall
conductance as the function of conduction electron density $n$ is
shown in Fig.~2(b), with the spin configuration fixed as Fig.~2(a).
The curve in Fig.~2(b) fluctuates dramatically, showing that the AHE
is sensitive to $n$ \cite{Takahashi:Prl}.

\begin{figure}
\vskip -1.8cm \centerline{\includegraphics[width=9cm]{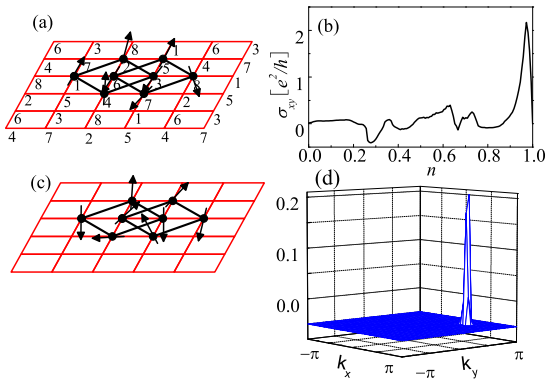}}
\vskip -0.7cm \caption{(Color online) (a) The ground state with the
parameter sets $J_1=J_2=0.04$, $J_3=0.025$, $t_1=1$, $t_2=0.25$,
$d=0.4$ and $n=0.5$. The unit cell expands to eight-site here. (b)
The Hall conductance as a function of the conduction electron
density $n$ with the spin order shown in (a). (c) An example of the
nontrivial spin order in the special eight-site unit cell. (d) Gauge
flux density $\textbf{B}(k)$ of the $1$-th band with the spin order
shown (c).} \vskip -0.5cm
\end{figure}

Interestingly, additional investigation of the ground state in
Fig.~2(a) indicates that besides the lattice distortion, this
special unit cell structure also breaks the P-symmetry
spontaneously. Thus for this unit cell structure, with the local
spin chirality, the band structure can lead to nonzero Chern number
and Hall conductance even without any lattice distortion or NNN
hopping. For the local spin order shown in Fig.~2(c), the eight
bands are topologically nontrivial and the corresponding Chern
number $C=[1, -1, 0, 0, 0, 0, -1, 1]$, with the gauge flux density
of the $1$-th band shown in Fig.~2(d). In fact, due to the
frustrated magnetic interaction, the unit cell could involve more
sites and become even larger with the parity violation satisfied.
For instance, for Eq.~(2), with large $J_1$ ($J_2$=$J_1$) and low
$n$, the variational calculation indicates the ground state is with
the coplanar spiral order. This coplanar spiral order can evolve
into the $3$-D spiral order by including the $H_2$ and $H_3$ term.
However, in this case, the unit cell also expands dramatically, and
the band calculation becomes much more challenging. The ground state
we address here seems to be a special case with a relatively small
unit cell.

Although the calculation of Eq.~(2) and the addressed two mechanisms
are more or less theoretically oriented, their realization in real
systems is still possible. For example, multiferroic TbMnO$_3$ is a
promising candidate to illustrate these two mechanisms and observe
the AHE. In TbMnO$_3$, besides the GeFeO$_3$-type distortion, which
forms the Mn-O-Mn zigzag chain, the DM interaction is also present
\cite{Arima:Prl}. In this case, due to the spiral order of the
background spin, the Mn-O-Mn zigzag chain is distorted with all the
oxygen ions displaced in the same direction \cite{Arima:Prl}.
Therefore the adjacent Mn-O-Mn angles are different, leading to the
staggered order of the effective Mn-Mn hopping amplitude and broken
parity symmetry. For TbMnO$_3$, neutron scattering experiments
confirmed that the background $t_{\rm {2g}}$ electrons form a
coplanar spiral order. To excite the noncoplanar spin order, a
magnetic field with its direction different from the spin order
plane can by applied, allowing the nonzero Hall conductance. In
addition, the ferroelectric polarization in TbMnO$_3$ aligns along
the direction of
$\textbf{e}_{ij}\times(\textbf{S}_i\times\textbf{S}_j)$. Therefore,
the noncollinear ferroelectric polarization should be expected for
this non-coplanar spin order. Moreover, the Tb$^{3+}$ cations can be
doped by other $+4$ cations, which can modulate the itinerant
electrons density and directly control the AHE \cite{Takahashi:Prl}.
However, for these calculations, a more practical two-orbital model
should be employed.

In conclusion, we have studied the Kondo lattice model on a square
lattice with frustrated super-exchange interactions. The calculated
nonzero Hall conductance is attributed to two distinct mechanisms,
one is the lattice distortion and the other is the locally
nontrivial spin order. Both of these mechanisms break the P-symmetry
and can generate the AHE spontaneously.

The authors acknowledge with E. Dagotto, Q. H. Wang, N. Nagaosa and
R. Shindou for fruitful discussions.


\begin{thebibliography}{bib}
\bibitem{supplement:none} In
the large $d$-limit, we take $f_1(k)\approx id\sin(k_x/2)$. This
approximation is not accurate especially at $k_x=0$. We take this
approximation here just to emphasize the importance of $t_{r2}$.
\end{thebibliography}
\end{document}